# Optomechanically and Themo-optically driven Interactions between Gilded Vaterite Nanoparticles in Bubbles

Hod Gilad[1,2*], Andrey Ushkov[1,2], Denis Kolchanov[1,2], Andrey Machnev[1,2], Toms Salgals[4], Vjačeslavs Bobrovs[4], Hani Barhum[1,2,3], and Pavel Ginzburg[1,2*]

[1]Department of Electrical Engineering, Tel Aviv University, Ramat Aviv, Tel Aviv 69978, Israel
[2]Light-Matter Interaction Centre, Tel Aviv University, Tel Aviv, 69978, Israel
[3]Triangle Regional Research and Development Center, Kfar Qara' 3007500, Israel
[4]Institute of Telecommunications Riga Technical University Azenes Street 12, Riga 1048, Latvia

**Abstract**

The capability to tailor mutual interactions between colloidal nanoparticles strongly depends on the length scales involved. While electrostatic and optomechanically driven interactions can cover nano and micron-scale landscapes, controlling inter-particle dynamics at larger distances remains a challenge. Small physical and electromagnetic cross-sections of nanoparticles make long-range interactions, screened by a fluid environment, inefficient. To bypass the limitations, we demonstrate that forming micron-scale bubbles around gilded vaterite nanoparticles enables mediating long-range interactions via thermo-optical forces. Femtosecond laser illumination leads to the encapsulation of light-absorbing particles inside long-lasting micron-scale bubbles, which in turn behave as negative lenses refracting incident light. Our experiments reveal the bubble-induced collimation of laser beams, traversing over mm-scale distances. The collimated beams are visualized with the aid of phase-contrast Schlieren imaging, which reveals refractive index variations, caused by temperature gradients within the fluid. We demonstrate that the refracted beams initiate the formation of secondary bubbles around nearby gilded vaterite particles. As the consequence, we demonstrate the ability to control secondary bubble motion by pushing and pulling it with optical radiation pressure force and by thermocapillary Marangoni effect, respectively. The latter facilitates interactions over millimeter-scale distances, which are otherwise unachievable. Apart from exploring new physical effects, mediating long-range interactions can find a use in a range of applications including drug design and screening, photochemistry, design of colloidal suspensions, and many others.

*Corresponding author

**Introduction**

Tailoring colloidal particle dynamics and laser particle interactions are important objectives across multiple disciplines, including fluid engineering, drug delivery systems, chemical catalysis, and many others [1–4]. Optical tweezers are the mastered apparatus for controlling the motion of micron-scale objects with the aid of laser beams. Since its invention [5], optomechanical trapping become a widely used tool in microbiological and biomedical studies [6–8]. However, admitting the complexity of interaction with a fluid environment, a range of different forces associated with non-isothermal flows can emerge [9]. In many cases, those can prevail over optomechanical interactions and govern the dynamics of micro-particles, immersed in a fluid. Management of thermal forces does not require optical transparency and, in many cases, can be employed in deep-tissue applications [10]. However, this comes at the expense of precision, which can be granted by focused laser beams. In this context, combining the advantages of optical and thermal forces inspires the investigation of controllable thermo-optical interactions. Hereinafter, we first discuss the particles, which will be used as a case study, and then bubble formation in the fluid will be revealed before introducing the interaction scenario.

Polycrystalline particles, capable of accommodating biochemical and optical properties within the same platform, were chosen for the studies. Within a range of inorganic particles [11,12], calcium carbonate ($CaCO_3$) has many advantages owing to its low-cost facile fabrication, biocompatibility, and biodegradability [13,14]. $CaCO_3$ has three crystalline forms - vaterite, calcite, and aragonite [15]. Vaterite undergoes a phase transition upon an interaction with a solvent and can dissociate or become calcite. In both of those scenarios, encapsulated functional materials will be released. For the sake of the experimental demonstration here, all experiments were done in ethanol (96%) to prevent vaterite recrystallization [16–18]. Apart from polymer encapsulation [19], the dissolution can be inhibited by positioning a particle in a bubble. However, pure vaterite is transparent material in visible and near-IR regions with an anisotropic refractive index of ~1.5-1.7 [20], which makes it ineffective as a candidate for laser heating applications due to its limited ability to absorb and retain heat efficiently. Consequently, introducing contrast and absorbing materials into the vaterite cargo can provide a solution to this problem. Decorating vaterite with plasmonic nanoparticles enables the tuning of Mie resonances, enhancing its potential as a functional biogenic metamaterial known as golden or gilded vaterite [21–23]. However, as it will become evident, nanoparticles, as they are, can also weakly interact with each other via optical fields thus demanding the development of auxiliary solutions to enlarge the action distances. Moreover, recent advancements have demonstrated that lasers can influence particles even beyond the immediate vicinity of the focal point, expanding the potential for remote manipulation in nanotechnology applications [24]. Here we propose using bubbles, controllably inflated around the particles, and this way to mitigate the interactions.

Themselves, gas bubbles in a solution possess various unique properties and can serve multiple functions, from medical diagnostics as agents for ultrasound imaging to 3d imaging and pharmacology [25–27]. This makes our basic investigations transformative.

To sustain micron-scale bubbles for extended periods without continuous energy input, one can use light-absorbing particles illuminated by a femtosecond laser source [28,29]. This approach differs from typical boiling, where a continuous energy supply balances vapor pressure and surface tension. Instead, short laser pulses heat the liquid, driving gas molecules previously dissolved in the solution to form bubbles. Remarkably, these bubbles can remain stable even after the laser is turned off. The durability of these bubbles does not rely on vapor condensation but on the slower process of gas redissolving into the liquid, which prevents immediate bubble collapse.

Our study case is to investigate thermo-optical interactions between gilded vaterite nanoparticles, encapsulated in bubbles (Fig. 1). To mitigate long-range particle-particle interactions, both optomechanical and thermal interaction scenarios will be explored, demonstrating the capability to control collective motion at relatively large millimeter-scale distances. Specifically, we will demonstrate that optically trapped long-living bubbles collimate laser beams and induce secondary interactions with other bubbles. Depending on the arrangement, both repulsion and attraction forces can be obtained. Despite an apparent complexity, all those scenarios are reproducible and stable.

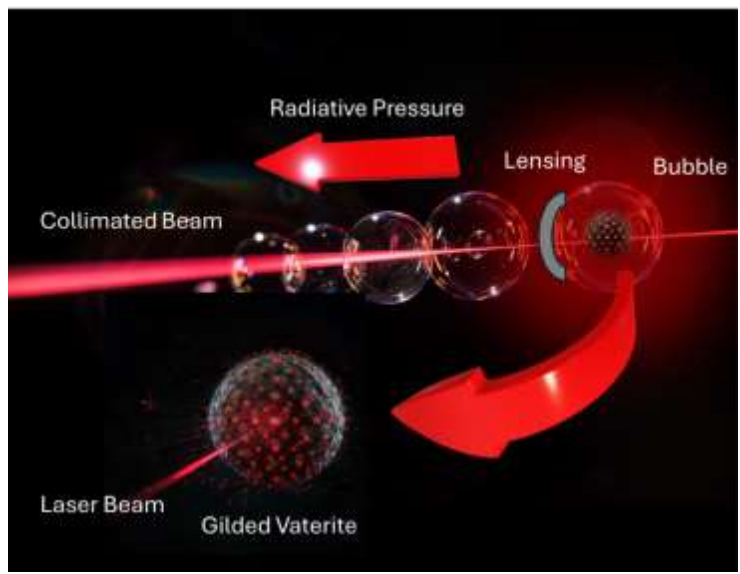

Figure 1: Illustration of the light-particle-bubble thermo-optical and optomechanical interactions.

**Methods**

**1. Synthesis of Vaterite Nanoparticles**

The synthesis of spherical vaterite particles is carried out using a precursor solution composed of 85% ethylene glycol and 15% aqueous solution, resulting in a final concentration of 0.05M CaCO3 and 0.01M $Na_2CO_3$. The mixture is stirred at 400rpm in a 50mL Erlenmeyer flask with a total volume of 40mL ethylene glycol. After 60 minutes of reaction, the mixture is centrifuged at 5,000rpm and washed once with deionized water (DIW) and twice with ethanol, subsequently being stored in an ethanol solution for further use.

## 2. Gilding of Vaterite Nanoparticles

The vaterite particles are coated with gold nanoparticles through a multi-step process [17]. Initially, an APTMS layer is added using the Stöber method, involving the mixture of 20μL APTMS, 50μL ammonia, and 930μL ethanol with 30-40mg of vaterite, and vortexed for 2 hours. After washing twice with ethanol, a PVP polymer solution (5mg/mL) at 5°C is added (2mL) and vortexed for another 2 hours. Post another ethanol wash, 3mg of vaterite is suspended in 4mL ethanol, to which 200μL of 25mM $HAuCl_4$ is added and mixed for 1 hour at 5 degrees Celsius. Triethylamine is then added to catalyze the reaction, which is left to proceed for 6 to 24 hours before collection. A typical particle's SEM image appears in Fig. 2(b).

## 3. Experimental Setup for Bubble Creation and Optical Measurement

### 3.1 Bubble Creation

To create long-living micron-scale bubbles, a pulsed laser source illuminates the gilded vaterite nanoparticles immersed in an ethanol solution. The heating with short laser pulses induces the formation of bubbles by extracting dissolved gas molecules from the solution, maintaining these bubbles long after the termination of light illumination. This mechanism diverges from traditional vapor condensation, relying instead on the slow re-dissolution of gas into the fluid [29,30]. A typical image of a bubble appears in Fig. 2(c).

### 3.2 Optical Setup

The experimental setup for probing particle interactions and observing bubble dynamics includes a 4f system with 100mm and 150mm lenses, a Schlieren filter in the Fourier space for temperature and refractive index variation visualization, and a FLIR Chameleon camera. Illumination is provided by a collimated halogen lamp. A MenloSystems YLMO-2W femtosecond laser, focused onto the sample using a Mitutoyo M-Plan X50 objective, heats the vaterite particles to initiate bubble formation. The setup also includes polarization optics and a polarizer, to control beam power, enabling the precise manipulation and heating of particles. The setup is depicted in Fig. 2(a), while a Schlieren image of a bubble and the collimated beam, heating the fluid, appears in Fig. 2(c) and will be elaborated in the 'Results' section.

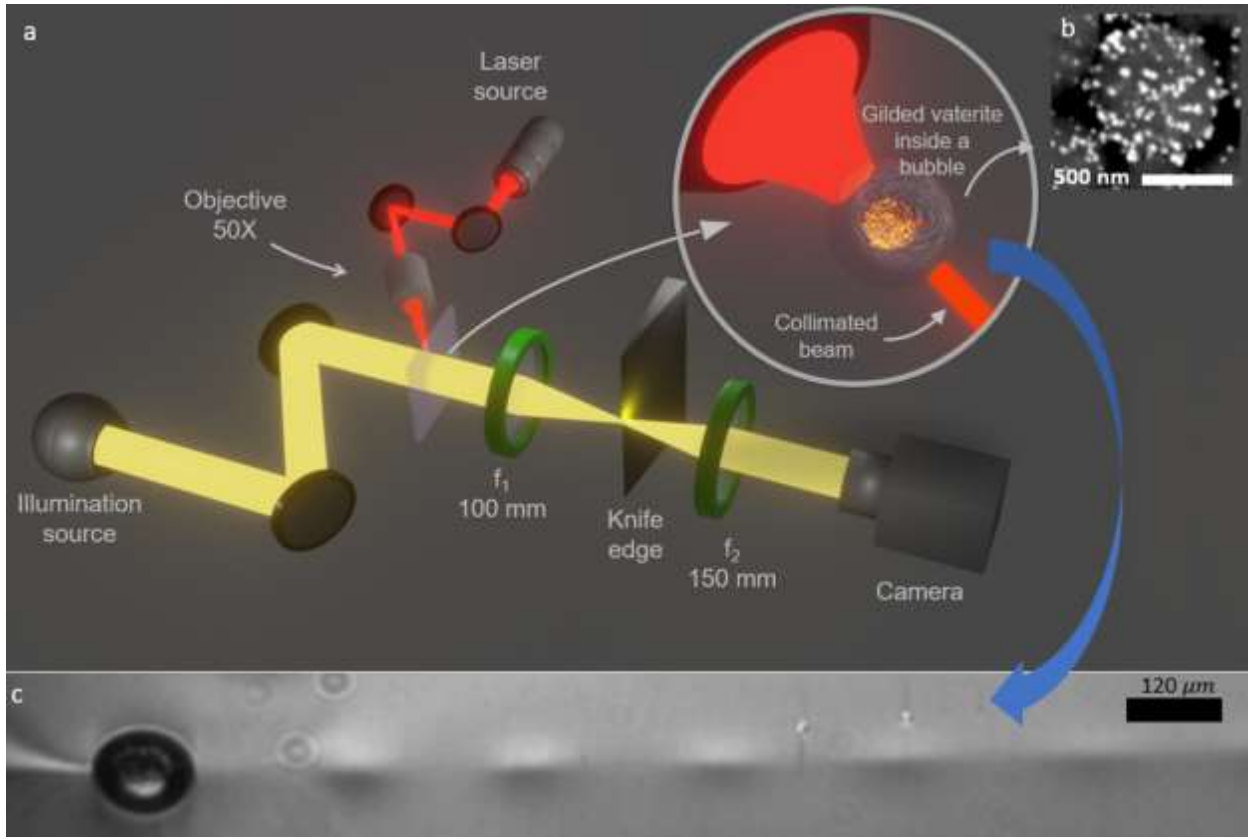

Figure 2. (a) Experimental setup of 4f Schlieren imaging and a femtosecond laser to create and manipulate bubbles within a solution. (b) SEM image of the gilded vaterite particle. (c) Schlieren's image of the bubble that was formed around the gilded vaterite after heating with the femtosecond laser.

## Results

*Bubble Creation*

Bubbles were created by illuminating gilded vaterite particles with the IR femtosecond laser, following the scheme in our recent report [29] (the experimental arrangement appears in Fig. 2(a)). Owing to strong light absorption, the gilded vaterite particle (SEM image in Fig. 2(b)), heats up and elevates the temperature of the embedding fluid – schematic appears in the inset to Fig 2(a). The bubble remained in the laser focus, experiencing optomechanical and thermo-optical attraction. It is worth noting that the thermo-optical forces prevail over optomechanical and are responsible for the bubbles' tweezing. However, the initial trapping of the particle has an optomechanical nature and, after bubbles were created, thermal gradients in the solution govern the dynamics. This statement, previously revealed in [29], will also become evident from the observations, presented hereafter.

*Bubble-induced Collimation*

Refractive index contrast between the gas bubble and the fluid environment (ethanol in this case) significantly changes the light propagation properties, creating a secondary lensing effect. Specifically, we observed that the trapping laser initially focused on a particle, interacts with a bubble, and is thus transformed into a collimated beam with ~60 $\mu m$ waist. Calculations, performed hereinafter, demonstrate that its intensity remains relatively high ~ $1.35 \cdot 10^8 \frac{W}{m^2}$. This new collimated beam was found to interact with other particles, creating secondary bubbles around them. To re-emphasize, the primary focused laser beam interacts with a single particle, and the emergence of other bubbles is unambiguously attributed to the light refraction. Several peculiar interactions between bubbles, containing gilded vaterite nanoparticles were observed and will be analyzed in detail. Before that, light refraction on bubbles and collimated beam formation is analyzed.

As a mere approximation, a bubble can be considered as a sphere, immersed within a high refractive index fluid. According to Lensmaker's equation, the focal length ($f$) of this structure can be calculated by [31]:

$$\frac{1}{f} = \left(\frac{n_l}{n_m} - 1\right)\left(\frac{1}{R_1} - \frac{1}{R_2}\right) = 2R\left(\frac{n_l}{n_m} - 1\right), \tag{1}$$

where $n_l$ is the refractive index of the lens (~1 in our case since the bubble is filled with gas at a relatively low pressure) and $n_m$ ~ 1.36 is the refractive index of the surrounding medium (ethanol). Since $n_l < n_m$, the structure acts as a negative lens, e.g., [4,32]. The spherical shape of the bubble dictates $R_1 = -R_2 = R$. In the case of a convergent incident beam, the negative lens can act as a collimator. To reveal this phenomenon, ray tracing modeling has been performed. Fig. 3 summarizes the numerical results, highlighting that both the radius of the bubble and the location of the beam waist with respect to the bubble's center (the shift) play a role. Fig. 3(a) is the schematics of the geometrical arrangement. Fig. 3(b) shows the beam spread angle (in degrees) as a function of the bubble's radius, for different bubble locations with respect to the beam waist. For bubbles with radii larger than 50 µm, the beam divergence has a very weak dependence on the parameters, thus the system remains robust to fluctuations. In other words, the forthcoming experimental demonstrations can be predicted to provide repeatable results, using many different particles and realizations in a fluid environment. Fig. 3(c) demonstrates the spread angle as a function of bubble-to-waist distance, while the bubble radius is kept constant. 3 different radii were explored. A similar conclusion can be deduced – keeping the bubble radius above 50 µm allows for achieving long-range collimated (below 10° divergence angle) beams along the sample. In the experiments, the bubble radius was repeatably obtained around 100 $\mu m$ and the bubble was kept before the beam waist, i.e., a positive shift between the bubble and beam waist.

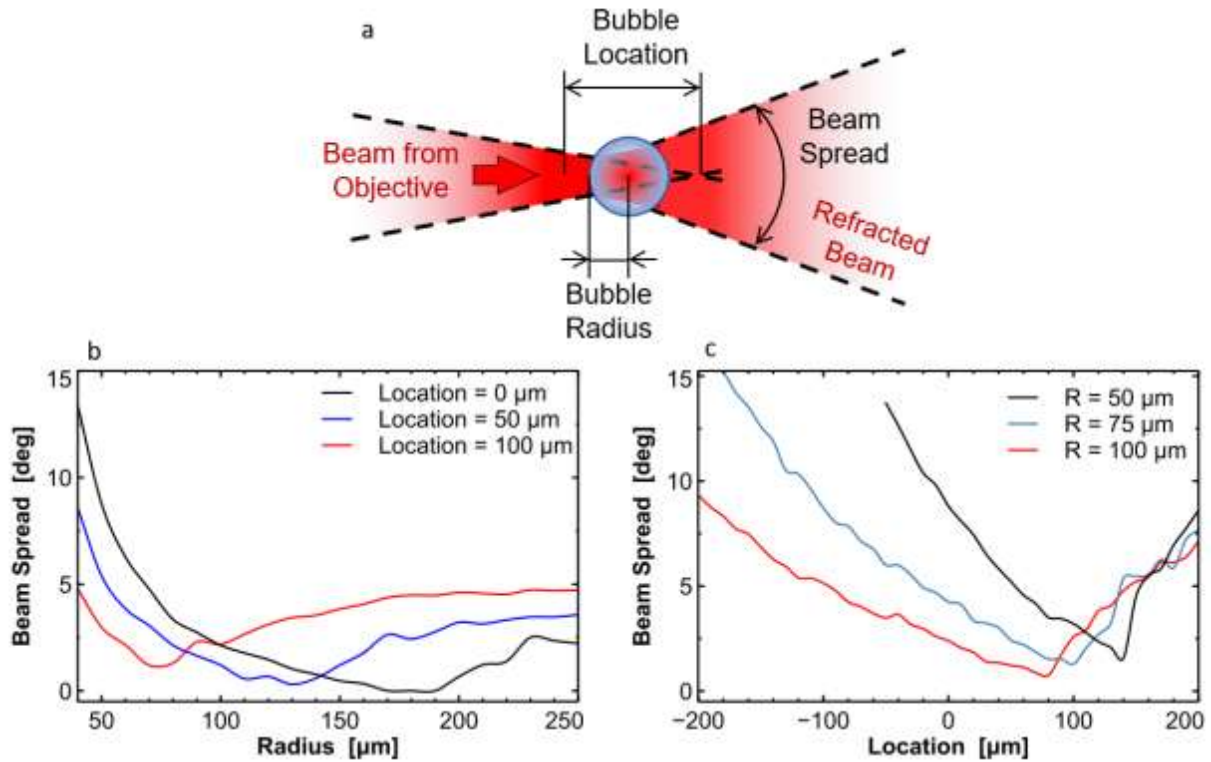

Figure 3. (a) The sketch of the considered geometry and the main parameters. (b) Beam spread as a function of the bubble's radius, for three different bubble positions with respect to the beam waist. (c) Beam spread as a function of bubble location with respect to beam waist, for three different bubbles radii.

*Schlieren imaging of bubble-collimated beams*

Direct observation of the beam, which does not encompass k-vectors, falling within an NA of collecting optics, is a challenging task. Scattering on particles, captured within the beam volume, enables reconstruction of the light intensity profile. While this method provides a solution for a free space (e.g., scattering on dust particles in air), it is less applicable in the case of pure liquids. Here, an alternative visualization technique, based on thermo-optical effects, is implemented. For example, Schlieren imaging can be utilized to indirectly capture changes in refractive indices in cases where material properties undergo variations in temperature or pressure [33,34]. Weakly absorbing light-transparent ethanol is one of the options, where thermo-optical effects can be captured.

Schlieren imaging is implemented on a 4f system and employs a razor to obstruct half of the beam at the k-space. While this filter leads to a 50% decrease in the intensity of the resulting image, it highlights the refractive index gradients. Visualization of bubbles and non-isothermal flows was reported recently in [35].

Worth noting the existence of many different methods for contrast imaging, e.g., phased contrast imaging and dark field imaging [36,37]. Schlieren is a reasonable compromise between the field of view, image quality, and ease of implementation, hence it was favored for use here. Schlieren highlights phase gradients normal to the knife edge, whereas phase-contrast methods are typically isotropic and indicate the phase change values.

Fig. 2(c) demonstrates the non-isothermal flow of liquid around the bubble and highlights the collimated beam. The beam has a very low divergence over 1.5 millimeters, falling to the field of view of the imaging system. To relate the visualized pattern with the temperature changes, a heat diffusion equation has been solved. A uniform light spot was assumed, and the absorption power was calculated by taking the imaginary part of ethanol's refractive index. The calculation shows that the local heat of a fluid of at least 6°C is possible within ~ 0.01 seconds. Note that this estimation is related to pure ethanol-light interaction. The gilded vaterite acts as a much more efficient heat source owing to its high absorption cross-section, boosted by the localized plasmon resonance phenomenon. A 6°C temperature increase reduces ethanol's refractive index by 0.0022 [38]. A contrast threshold of 5% is sufficient for a noticeable change in the image. An equation to relate Schlieren imaging transforms $\Delta n$ (change in refractive index) into a detectable brightness variation is given by [33]:

$$\frac{\partial n}{\partial x}_{min} = 0.05 \cdot \frac{n_0}{L} \cdot \frac{a}{f_2} \qquad (2)$$

where 0.05 is 5% which is the threshold value, and the other parameters in our experiment are $n_0 = 1.36$ (at room temperature, 22°C), L=400 $\mu m$ is the length through inhomogeneity, $a = 2$ mm is the radius of the beam in k-space between the 2 lenses, that is undisturbed by the knife edge as in Fig. 2(a), and $f_2 = 150$ mm is the focal length of the second lens. In our experiment $\frac{\partial n}{\partial x_{min}} \approx 45.33 \frac{1}{m}$, while the gradient value is $\frac{\partial n}{\partial x} \approx \frac{0.0022}{5\ \mu m} = 440 \frac{1}{m}$, which is 10 times larger than the minimal detectable value, and thus the collimated beam can be made visible as it is under the provided experimental conditions. It is worth noting that the collimated beam image (Fig.2(c)) disappeared after the razor was removed, validating the necessity of the phase-contrast approach.

*Secondary Bubble Generation*

The collimated beam, formed by a primary bubble, might encounter another gilded vaterite particle and, consequently, initialize the formation of the secondary bubble. Worth noting that without the collimation effect, the electromagnetic intensity on the secondary particle will be insufficient to create the bubble. While in the case of a static particle (initiator), the bubble growth dynamics, tailored by laser light, follow the model, reported in [28], the fluid environment can exhibit different scenarios. Fig 4(a) is the comparison between several bubble growth regimes when: (1) the initiating particle is localized within the laser focus (red dots – experiment, black solid line is the theoretical fit, based on [28]), (2) there is a fluid flows around the particle during the bubble growth (green crosses) (up to 150 $\frac{\mu m}{s}$ velocities were induced in the fluid cell), and (3) bubble growth occurs within the collimated beam (blue circles). Those three interaction scenarios are schematically summarized in Fig. 4(b, c, and d). Experimental data shows that all those regimes lead to completely different bubble growth dynamics. In the first scenario, G. Baffou et. al. argue that the deceleration of the growth rate is attributed to the depletion of gas molecules around the bubble [28]. Following this argument, the constant liquid flow is expected (a) to provide an undeleted supply of gas molecules, resulting in a faster growth rate, and (b) to reduce the local temperature owing to the heat removal, slowing down the bubble inflation. The experimental data (green crosses) indeed demonstrates the linear growth rate though with a smaller slope, compared to the first scenario. The third interaction type further supports this claim. The collimated beam and the fluid flow directions are co-aligned, thus the undeleted gas concentration and higher local temperature along the beam result in a much faster growth rate. The bubble grew much faster in the experiment until its size exceeded the collimated beam diameter. All videos for all the regimes appear in the Supplementary Information.

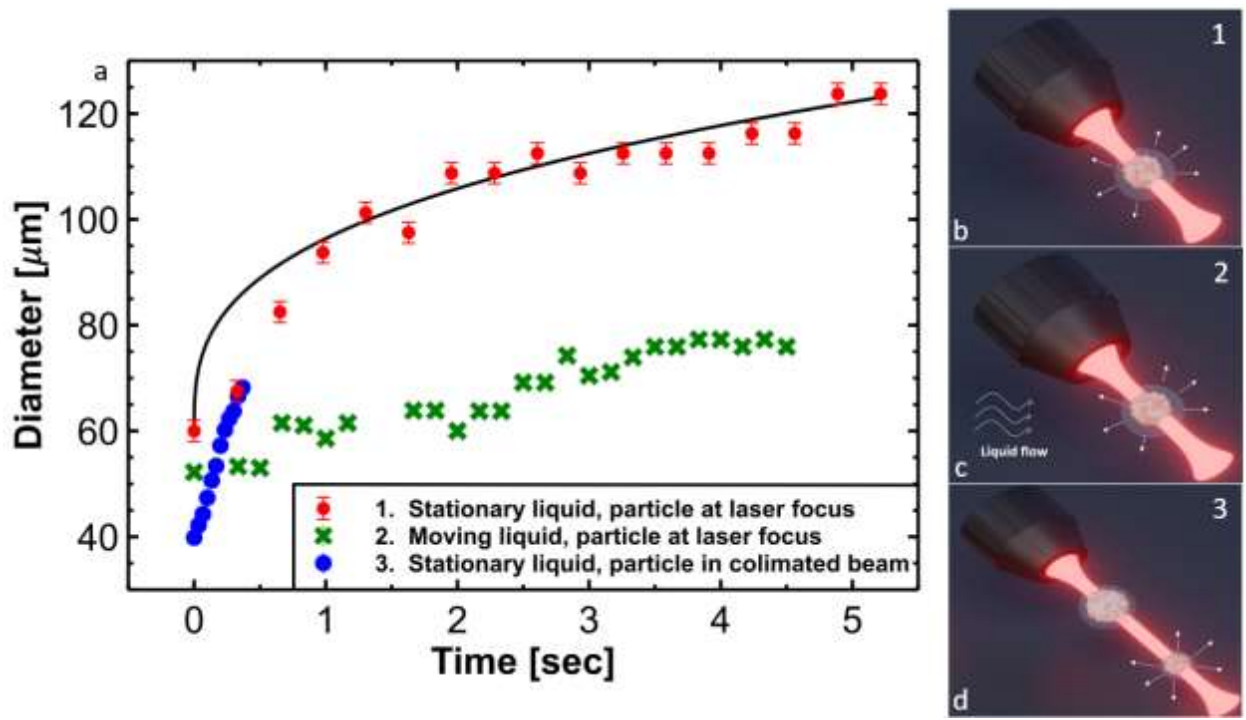

Figure 4. Bubble growth dynamics around gilded vaterite nanoparticle, illuminated with a femtosecond laser beam. (a) Experimental data and theoretical fit (as in the legend). (b-d) Different interaction regimes: (b) The bubble and the particle are trapped in the laser focus, and the fluid (ethanol) is stationary.

(c) The bubble and the particle are trapped in the laser focus, and the fluid is driven into the constant flow. (d) The bubble and the particle are within the collimated beam, which drives them into the synchronized flow.

*Bubble-bubble repulsion*

To initiate light-mediated bubble-bubble repulsion, the following scenario has been explored. Several bubbles were created first with the aid of the femtosecond laser, which was switched off at the end of this action. Recall that bubbles, encompassing the gas molecules and not the vapor, remain stable over time (minutes and hours) after the light is switched off (a scheme appears in Fig. 5(a)). In the second stage, after the liquid is cooled down and the temperature gradients vanish, the laser is switched back again. After interacting with a bubble, the laser undergoes collimation (as elaborated before) and, within a certain probability, encounters a secondary bubble applying radiation pressure on it (a scheme appears in Fig. 5(b)). Since the size of the bubble is significantly larger than the wavelength, the principles of geometrical optics apply, leading to [39]:

$$F_{ref} = \frac{2R_f P}{c},  \qquad (3)$$

where $R_f = 0.0228$ is the Fresnel reflection coefficient (for intensity), calculated for the ethanol-air interface: $n_e = 1.356, n_a = 1$, $c$ is the speed of light, and $P$ is the beam power. Substituting the experimentally relevant values, i.e., $P = 1.35 \cdot 10^8 \cdot \pi \cdot (40 \mu m)^2 = 0.93\ W$, results in $F_{push} \sim 150\ pN$. Recall that the radiation pressure on a single gilded vaterite nanoparticle will be negligible without the embedding bubble. The bubble motion dynamics as a function of time are summarized in Fig. 5(c). The bubble velocity increases with time, as long as it experiences the acceleration from radiation pressure force. When the collimated beam is off (the laser is shut down), the bubble slows down till it stops owing to friction. A corresponding video is in the Supplementary Information. Fig. 5(d) shows several video frames, highlighting the dynamics.

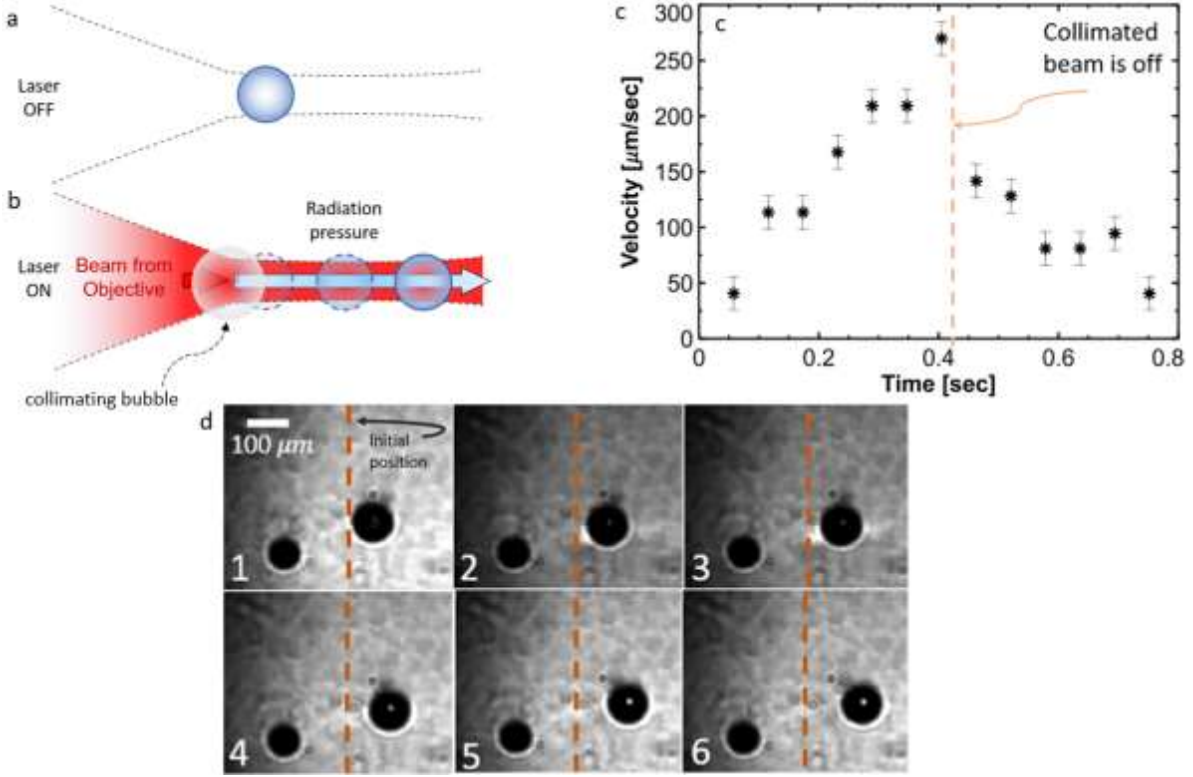

Figure 5. Bubble-bubble repulsion. (a, b) A sketch of the experiment, showing stable bubbles without a laser source. When the laser is 'on', it is collimated by a primary bubble and pushes the secondary with the radiation pressure. (c) Secondary bubble velocity as a function of time. (d) The bubble dynamics – propulsion of the secondary by the collimated beam. The dashed line shows the initial location. Times are: 0, 0.17, 0.29, 0.35, 0.46, 0.58, 0.69 seconds, respectively. A burst of the collimated beam can be seen in pictures number 2 and 3 and then disappeared in others.

*Secondary bubble attraction – thermocapillary effect*

The thermocapillary effect is known by pulling non-liquid materials into areas in liquid where the surface tension is lower [40]. In ethanol, as with most liquids, the surface tension gets lower when the temperature increases. Therefore, since the collimated beam heats the liquid (Fig. 2(c)), it creates temperature gradients, causing bubbles to be drawn toward the beam (video in the Supplementary Information). A similar effect was observed with a convergent beam in our previous study [29]. As the S3 video shows, when the collimated beam was off, buoyancy force controls bubble motion. The relevant force is $f_B = \frac{4\pi R_b^3}{3} \cdot g \cdot \rho_l \approx 2.07 \cdot nN$, when $R_b = 40\ \mu m$ is bubble radius, $g = 9.8 \frac{m}{s^2}$, and $\rho_l = 789 \frac{kg}{m^3}$ is ethanol density. When the collimated beam appears, it induces a thermocapillary force, which is estimated as follows: $F_{th} = -2\pi R_b^2 \nabla T \frac{d\sigma}{dT} \approx 646\ nN$, where $R_b = 40\ \mu m$ is the bubble radius. According to numerical simulation

$\nabla T = 6^o$ between the liquid inside the collimated beam and the distant background. According to [41], $\frac{d\sigma}{dT} \approx -0.0009 \frac{N}{m \cdot k}$, resulting in $F_{th} \approx 646\ nN$, which is significantly stronger than the buoyancy force. Optical force and gravity are negligible here. The dynamics are graphically summarized in Fig. 6(a, b). Fig. 6(c) demonstrates the secondary bubble velocity before and after interacting with the collimated beam. The transition between the bouncy and thermocapillary attraction is evident alongside the acceleration along the temperature gradient. Linear acceleration supports the claim on the linear temperature gradient.

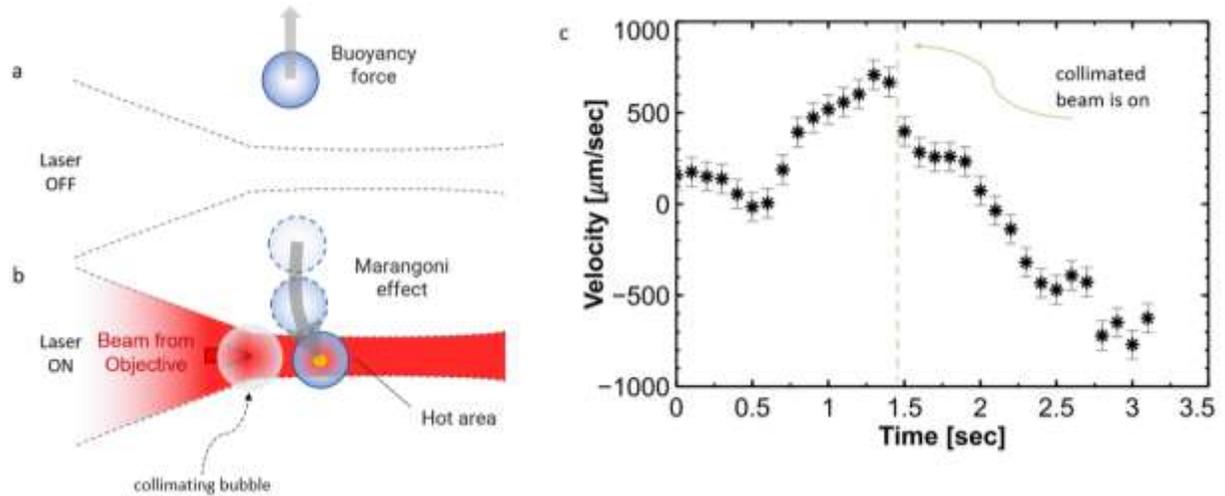

Figure 6. Secondary bubble attraction. (a), (b) Sketch of the bubble motion without collimated beam. (c) Bubble velocity, as the function of time - when the laser is off since buoyancy force governs the motion. At $t$=1.45 sec (green dashed line), the collimated beam is 'on' and the bubble is pulled back by the thermocapillary force to the hottest location, around a gilded vaterite.

**Conclusion**

The dynamic interactions between gilded vaterite nanoparticles in bubbles under the influence of optomechanically and thermo-optically driven forces were explored. Utilizing femtosecond laser illumination, we successfully encapsulated light-absorbing nanoparticles within micron-scale bubbles, serving as novel, negative lenses to refract incident light and create high-intensity, collimated beams over millimeter-scale distances. Our approach demonstrates the capability of controlling long-range inter-particle dynamics, overcoming the limitations posed by nanoparticles' small physical and electromagnetic cross-sections in a fluid environment, and showing the significant role of temperature-dependent surface tension and radiation pressure in mediating interactions. Applying controlled long-range interactions can find a use in various fields, including drug design, photochemistry, and the design of colloidal suspensions.

**Conflict of interest**

The authors declare no conflicts of interest.

**Acknowledgments**

This research was funded by the ERC StG "In Motion" (802279), Israel Science Foundation (ISF grant number 1115/23). A.U. acknowledges the support of the Azrieli Foundation's Postdoctoral Fellowship, and RTU Team acknowledges the Latvian Council of Science project: No. lzp-2022/1-0579.**References:**